\RequirePackage{tikz}

\documentclass[doublecol]{epl2} 

\usepackage{tikz}

\usepackage{amsmath}
\usepackage{amssymb}
\usepackage{bm}
\usepackage{verbatim}
\usepackage{graphicx}
\usepackage[lofdepth,lotdepth]{subfig}

\usepackage{hyperref}

\usepackage{lpic}
\usepackage{psfrag}
\usepackage{epsfig}
\usepackage{mathrsfs}

\newcommand{\gb}{\bm{\hat{g}}}

\newcommand{\eq}[1]{\begin{align}#1\end{align}}

\newcommand{\OO}{\mathcal{O}}

\newcommand{\ffrac}[2]{\mbox{$\frac{#1}{#2}$}}
\def\half{\mbox{$\frac{1}{2}$}}

\def\p{\partial}
\def\zbar{\bar{z}}

\def\sigmab{\bm{\hat{\sigma}}}

\def\rhob{\bm{\rho}}

\def\etab{\bm{\eta}}

\def\rb{\bm{r}}
\def\fb{\bm{f}}
\def\lb{\bm{\ell}}

\def\sb{\bm{s}}
\def\tb{\bm{t}}

\def\epsb{\bm{\hat{\varepsilon}}}

\def\delb{\bm{\hat{\delta}}}

\def\fgc{\bm{f}_g^c}

\def\tr{\mbox{tr}}

\title{On the Granular Stress-Geometry Equation}
\author{Eric DeGiuli\inst{1,2} \and Christian Schoof\inst{3}}
\shortauthor{Eric DeGiuli \etal}

\institute{                    
  \inst{1} Department of Mathematics, University of British Columbia, 
Vancouver, BC, Canada \\
  \inst{2} Center for Soft Matter Research, New York University, New York, NY, USA \\
  \inst{3} Department of Earth, Ocean, and Atmospheric Sciences, University of British Columbia, Vancouver, BC, Canada
}
\abstract{
Using discrete calculus, we derive the missing stress-geometry equation for rigid granular materials in two dimensions, in the mean-field approximation. We show that (i) the equation imposes that the voids cannot carry stress, (ii) stress transmission is generically elliptic and has a quantitative relation to anisotropic elasticity, and (iii) the packing fabric plays an essential role. 
}
\begin{document}
\maketitle
\section{\label{sect:intro}Introduction}

\newcommand{\gammab}{\bm{\gamma}}
\newcommand{\Gammab}{\bm{\hat{\Gamma}}}
\renewcommand{\etab}{\bm{\eta}}
\newcommand{\zetab}{\bm{\hat{\zeta}}}
\newcommand{\betav}{\bm{\beta}}
\newcommand{\betab}{\bm{\hat{\beta}}}
\newcommand{\lambdab}{\bm{\hat{\lambda}}}
\newcommand{\xib}{\bm{\xi}}
\newcommand{\Fb}{\bm{\hat{F}}}
\newcommand{\Gb}{\bm{\hat{G}}}

Despite a century of study, the macroscopic behaviour of quasistatic granular materials remains poorly understood \cite{deGennes:1999,RouxCombe:2010}. We still lack a fundamental system of continuum equations, comparable to Navier-Stokes for a Newtonian fluid. Experiments and simulations indicate that stress distribution within a granular solid depends on the packing's preparation history \cite{Vaneletal:1999,Sereroetal:2001,Gengetal:2001b,Gengetal:2003,Atmanetal:2005}. The latter is known to induce anisotropy in the statistics of grain arrangement, known as the packing fabric, characterized most simply by a 2nd-order symmetric tensor \cite{Oda:1972a,Oda:1972b,Calvettietal:1997,Gengetal:2001b,Gengetal:2003,Atmanetal:2005}. Hence the fabric, along with its evolution equation, may be the crucial internal variable needed to close macroscopic equations for quasistatic granular materials. 

Using tools of discrete calculus, in this work we derive one of the missing continuum equations for two-dimensional (2D) granular materials directly from the grain scale, in the mean-field approximation. The stress-geometry equation thus derived relates the stress tensor to the fabric. It shows that stress transmission in frictional granular materials is generically described by an elliptic equation, closely related to anisotropic elasticity, as previously suggested \cite{Sereroetal:2001,Bouchaudetal:2001,Ottoetal:2003}.

We consider isostatic packings of perfectly frictional, rigid grains in 2D, in the absence of gravity. The granular material is composed of $N_{RG}$ convex grains in mechanical equilibrium, touching at $N_{C}$ point contacts\footnote{Of the $N$ total grains in a real packing, a fraction $x_0$ will be geometrically trapped, but not contribute to mechanical stability. These `rattlers' are excluded from the $N_{RG}=N(1-x_0)$ force-bearing grains which are the subject of analysis.}. Mechanical equilibrium requires that
\eq{ \label{Newton}
0 = \sum_{c\in C^g} \fgc, \quad \;\;\; 0 = \sum_{c\in C^g} (\rb^c - \rb^g) \times \fgc,
}
where $\fgc$ is the contact force exerted on grain $g$ at contact $c$, $\rb^c$ is the position of $c$, $\rb^g$ is position of the centre-of-mass of $g$, and $C^g$ is the set of contacts belonging to $g$. Here the cross-product is defined as $\bm{a} \times \bm{b} = \bm{a} \cdot \epsb \cdot \bm{b} = a_i \varepsilon_{ij} b_j$ with $\varepsilon_{12}=-\varepsilon_{21}=1, \varepsilon_{11}=\varepsilon_{22}=0$. In what follows, all tensor contractions are explicitly indicated by a dot `$\cdot$'.

Newton's laws \eqref{Newton} give $3N_{RG}$ scalar constraints on the $2N_C=N_{RG}\zbar$ degrees-of-freedom in the contact forces, defining the contact number $\zbar$. When $\zbar=3$, the packing is \textit{isostatic}: given the positions and orientations of the grains and the external loading, Newton's laws can be \textit{solved} for the contact forces \cite{EdwardsGrinev:1999,Roux:2000}. 

The macroscopic object of interest is the stress tensor \cite{KruytRothenburg:1996} 
\eq{
\label{sigma}
\sigmab (\rb) = -\frac{1}{A_G} \sum_{g\in G} \sum_{c\in C^g} (\rb^c - \rb^g) \fgc,
}
where $G=G(\rb)$ is a set of grains centered on the point $\rb$, occupying the area $A_G$.  

In principle, the microscopic isostatic contact force solution can be coarse-grained to produce the macroscopic stress tensor. However, this is both computationally and analytically intractable. It would be preferable to determine the macroscopic $\sigmab$ by the solution to continuum equations. Mechanical equilibrium requires that the stress tensor satisfies \cite{ChaikinLubensky, DeGiuliMcElwaine:2011}
\eq{
\label{Newton_cont}
0 = \nabla \cdot \sigmab, \quad \;\;\; \sigmab = \sigmab^T,
}
however these 3 equations are insufficient to determine the 4 components of $\sigmab$. In passing from the grain scale to the continuum, one macroscopic equation has gone missing: the stress-geometry equation \cite{EdwardsGrinev:1999,EdwardsGrinev:2001,BallBlumenfeld:2002,Blumenfeld:2004}. 


A continuum equation can only apply when there is a large separation between microscopic and macroscopic length scales. The former is given by the grain scale, for example the mean grain diameter which we set to $1$; the latter, denoted, by $L$, arises from macroscopic boundary conditions, such as the domain size. Throughout we assume that $1/L \ll 1$.

Since contacts lack a preferred orientation and gravity is neglected, the stress-geometry equation must contain fabric tensors only of even order; the most general such continuum differential equation which is linear in stress is
\eq{ \label{eqseries}
0 & = \Fb_2 : \sigmab + \nabla \nabla : \big( \Fb_4 : \sigmab \big)+ \ldots
}
where $\Fb_n$ is an $n^{th}$ order fabric tensor, and `$:$' indicates 2 tensor contractions. On dimensional grounds, since fabric is microscopic, we expect that $\Fb_n \sim 1$ and $\nabla \sim L^{-1}$, so that higher terms are suppressed by powers of $1/L$. If $\Fb_2 \neq 0$, then \eqref{Newton_cont} and \eqref{eqseries} generically lead to a hyperbolic problem for $\sigmab$, and stress is transmitted along preferred directions; the Fixed Principal Axes and Mohr-Coulomb closures are of this type \cite{EdwardsGrinev:1999, Bouchaudetal:2001, BallBlumenfeld:2002}. However, if $\Fb_2 \equiv 0$, then \eqref{Newton_cont} and \eqref{eqseries} generically lead to an elliptic problem for $\sigmab$, and stress transmission would be more closely related to anisotropic elasticity. In this work, we resolve the hyperbolic/elliptic debate by deriving \eqref{eqseries} in the continuum limit, in the mean-field approximation, with explicit expression for the fabric tensors. 

To find the hidden equation, we make essential use of the voids in between the grains which, in two dimensions, are uniquely associated with closed loops of grains \cite{Satake:1992, BallBlumenfeld:2002}. We will show that the stress-geometry equation bears a simple physical interpretation: the voids cannot carry any stress. 

\section{Stress Potentials}

\begin{figure*}[ht!]
\centering
\begin{minipage}{0.35\textwidth}%
\subfloat
{
\begin{lpic}[nofigure]{"EllipticalGrains2"(2.2in,3in)}
\includegraphics[viewport=500 50 1040 688,width=0.95\textwidth,clip]{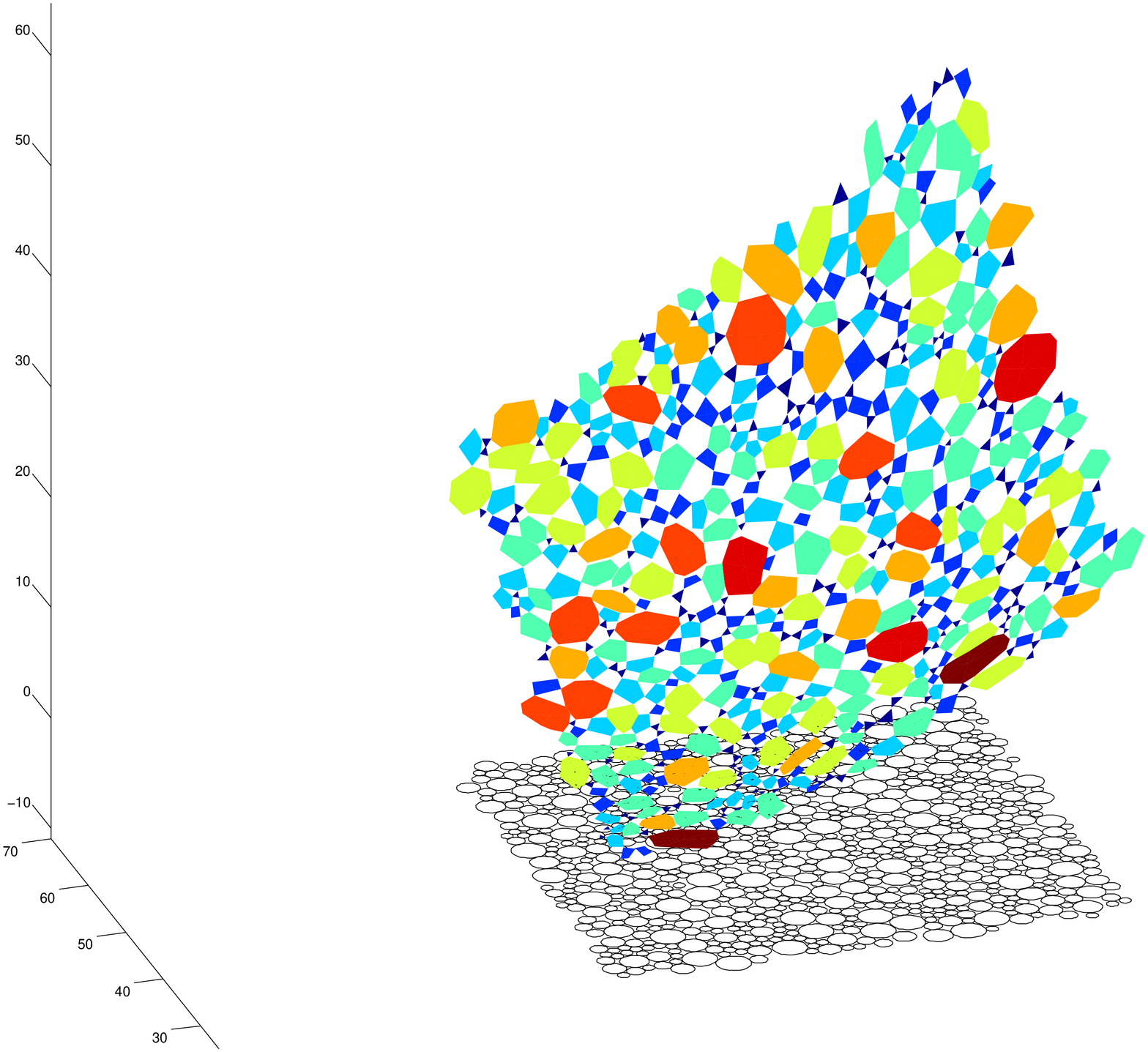}
\lbl[bl]{-220,10,0;(a)}
\end{lpic}
\label{fig:psi}}
\end{minipage}%
\quad
\begin{minipage}{0.62\textwidth}%
\centering
\subfloat
{
\begin{lpic}[nofigure]{"EllipticalGrains2"(3.8in,2in)}
\includegraphics[viewport=70 210 530 475,width=3.5in,height=2in,clip]{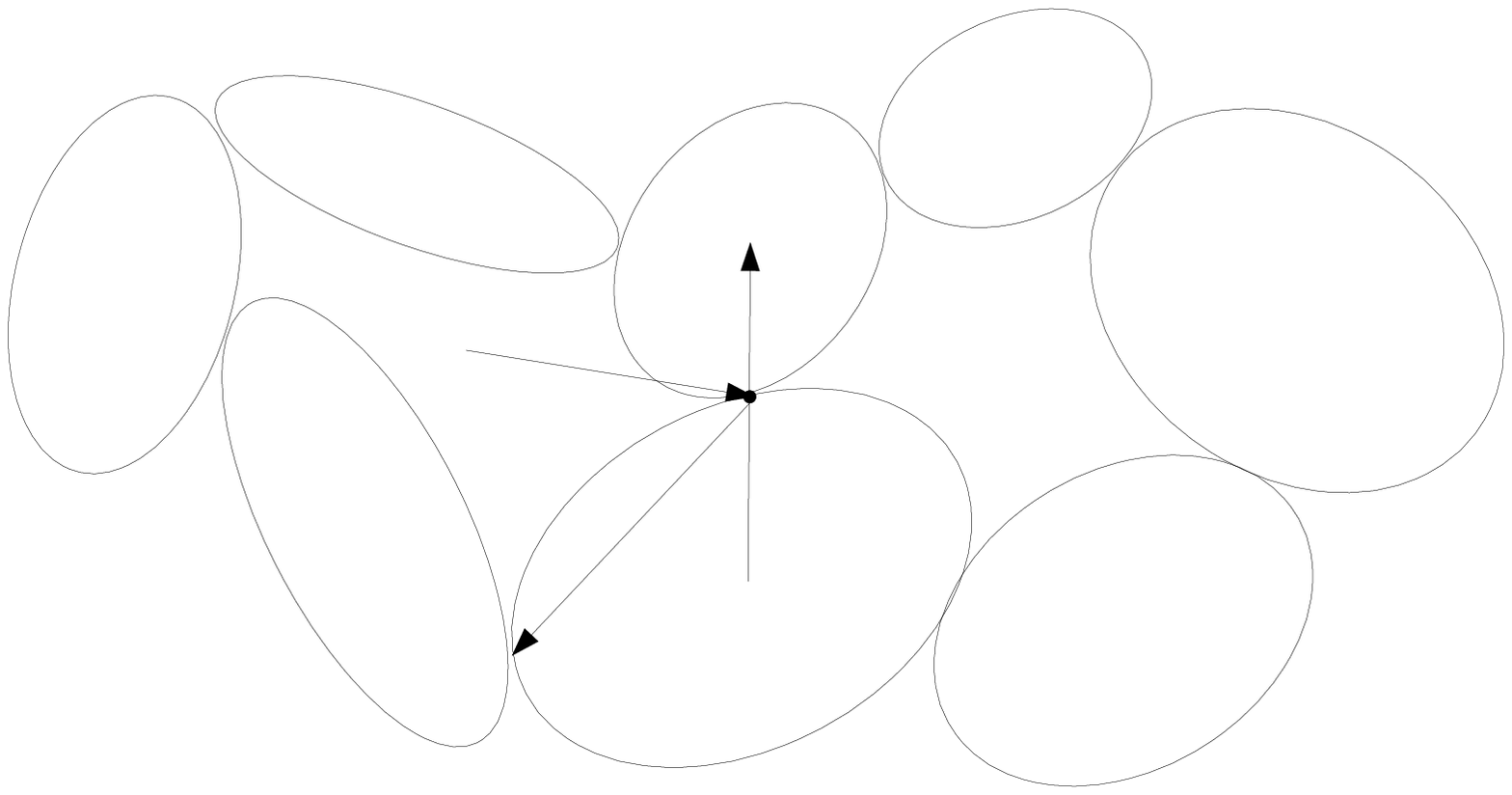}
   \lbl[bl]{-170,30,0;(b)}
   \lbl[bl]{-88,115,0;$c$}
  \lbl[bl]{-90,60,0;$g$}
  \lbl[bl]{-65,135,0;$\ell'$}
  \lbl[bl]{-130,140,0;$\ell$}
   \lbl[bl]{-88,170,0;$g'$}
   \lbl[bl]{-88,90,0;$\bm{\ell}^c$}
   \lbl[bl]{-108,65,0;$\bm{s}^\ell_g$}
   \lbl[bl]{-110,118,0;$\bm{t}^c_\ell$}
\end{lpic} 
\label{fig:notation}} \\
\subfloat
{
\begin{lpic}[nofigure]{"PsiSurfaceCrossSection"(3.8in,0.9in)}
\includegraphics[viewport=50 320 430 425,width=3in,clip]{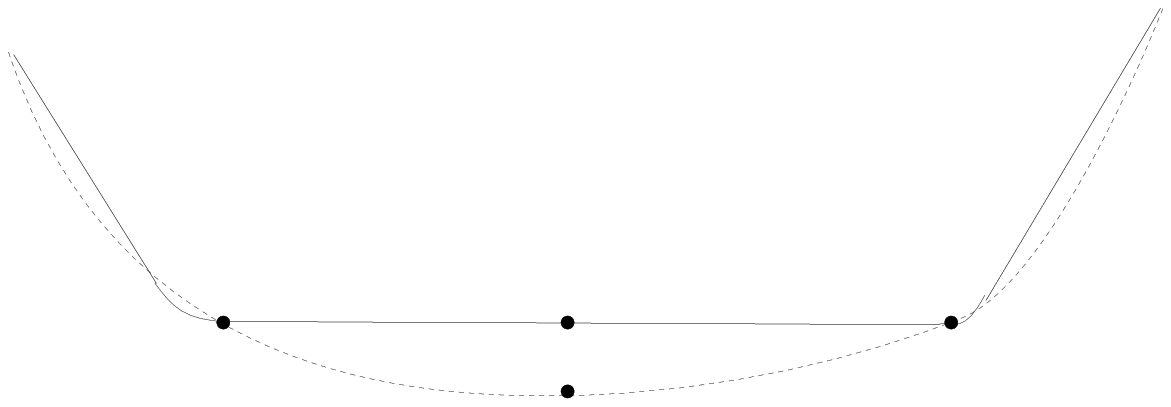}
  \lbl[bl]{-145,47,0;(c)}
  \lbl[bl]{-80,145,0;$\ell$}
  \lbl[bl]{-105,148,0;$\psi^{c_1}$}
  \lbl[bl]{-27,148,0;$\psi^{c_2}$}
  \lbl[bl]{-65,47,0;$\langle \psi \rangle(\rb^\ell)$}
  \lbl[bl]{-65,150,0;$\varphi^\ell$}
\end{lpic} 
\label{fig:cross}}
\end{minipage}
\caption{(a) Discrete $\psi$ surface. Each facet corresponds to a loop; colours correspond to the number of contacts around a loop. (b) Local geometry around contact $c$. (c) Cross-section of $\psi$ surface, in its original $\psi$ (solid) and smoothed $\langle\psi \rangle$ (dashed) versions. The contacts $c_1$ and $c_2$ are part of the loop $\ell$.}
\label{fig:globfig}
\end{figure*}

We define potentials $\rhob$ and $\varphi$ such that contact forces $\fgc$ and torques $\rb^c \times \fgc$ are written as
\eq{
\label{rhol1}
\fgc & = \rhob^{\ell'} - \rhob^\ell, \\
\rb^c \times \fgc & = \varphi^{\ell'} - \varphi^\ell + \rb^{\ell'} \times \rhob^{\ell'} - \rb^\ell \times \rhob^\ell, \label{varphil1}
}
where $\ell'=\ell'(c)$ ($\ell=\ell(c))$ is the loop to the right of (to the left of) the oriented contact $c$ (see Figure \ref{fig:notation}), and $\rb^\ell$ is the center  of loop $\ell$, defined below \cite{Satake:1993,DeGiuliMcElwaine:2011}. Writing the contact forces and torques in this way, force and torque balance are identically satisfied, for \textit{any} choice of $\rhob$ and $\varphi$. Conversely, the latter equations are precisely the conditions needed to write \eqref{rhol1} and \eqref{varphil1}. 
Given $\fb$, the potentials are unique up to an irrelevant additive constant. Equations \eqref{rhol1} and \eqref{varphil1} were first written down by Satake \cite{Satake:1993}. The formulation which uses $\rhob$ but not $\varphi$ was considered by Ball and Blumenfeld \cite{BallBlumenfeld:2002}.

The potentials are not independent, since the torques computed from $\rhob$ must equal those computed from $\varphi$. Writing $\tb^c_\ell=\rb^c-\rb^\ell$, this imposes $N_{C}$ constraints
\eq{ \label{consist}
\varphi^{\ell'} - \tb^c_{\ell'} \times \rhob^{\ell'} = \varphi^{\ell} - \tb^c_\ell \times \rhob^{\ell}.
}
Since all other constraints have been satisfied, \eqref{consist} are the discrete stress-geometry equations in this formulation. Our goal is to rewrite these equations in such a way that a continuum limit may be taken. Continuum expressions are naturally related to sums of discrete expressions around closed contours \cite{DeGiuliMcElwaine:2011}. Since, by Euler's formula, $N_{C} = N_L + N_{RG}-1$, with $N_L$ the number of loops, it is natural to sum these equations around grains and loops to form an equivalent set of constraints that are more easily interpreted as continuum equations. 

For example, summing \eqref{consist} around a grain, we find
\eq{ 
0 = \sum_{\ell \in L^g} \sb^\ell_g \times \rhob^\ell, 
}
where $\sb^\ell_g$ circulates anticlockwise around the grain (Figure \ref{fig:notation}). This expression can be rewritten as
\eq{ \label{divrho}
\sum_{\ell \in L^g} \sb^\ell_g \times \rhob^\ell \equiv \oint_{\p g} d\rb \times \rhob \equiv - A^g (\nabla \cdot \rhob)^g,
}
where the equivalences are definitions in discrete calculus \cite{DeGiuliMcElwaine:2011}, and $A^g$ is the area of the polygon formed by the $\tb^c_\ell$ vectors around $g$. 
To obtain continuum equations, we define area-weighted averaging operators
 \eq{ \label{average}
 \langle  P \rangle (\rb) = \frac{1}{A_G} \sum_{g \in G} A^{g} P^{g}, \;\; \langle  Q \rangle (\rb) =\frac{1}{A_L} \sum_{\ell \in L} A^{\ell} Q^{\ell}, 
}
for fields defined on grains and loops, respectively. The sets $G$ and $L$ are neighbourhoods around $\rb$, which must become arbitrarily large in the continuum limit. We suppose that each neighbourhood contains $M \gg 1$ grains. 

Upon averaging, $(\nabla \cdot \rhob)^g$ converges to its continuum counterpart in the following sense: a smooth function $\rhob(\rb)$ can always be defined such that $(\nabla \cdot \rhob)^g - \nabla \cdot \rhob(\rb^g) = \bm{\hat{\Gamma}}^g : \nabla \rhob(\rb^g) + \OO( \nabla \nabla \rhob(\rb^g))$, with $\bm{\hat{\Gamma}}^g$ a fluctuating fabric tensor with zero average\footnote{$A^g \Gammab^g =-A^g \delb - \sum_{\ell \in L^g} \rb^\ell s^\ell_g \cdot \epsb = \bm{\hat{P}}^g \cdot \epsb$ in terms of the fabric tensor $\bm{\hat{P}}^g$ considered by Ball and Blumenfeld \cite{BallBlumenfeld:2002}, with $\delb$ the identity tensor.}. This follows by Taylor expanding $\rhob(\rb^\ell) = \rhob(\rb^g) + (\rb^\ell-\rb^g)\cdot \nabla \rhob(\rb^g) + \ldots$, and collecting terms in \eqref{divrho}. 
As shown previously \cite{BallBlumenfeld:2002}, the volume average of $A^g \Gammab^g$ cancels up to boundary terms, which form a contour around the averaging area, of length $\sim\!\! \sqrt{M}$ grains. If the fabric has a finite correlation length $\xi_f$, then its average over $M$ grains is composed of $\sim\!\sqrt{M}/\xi_f$ independent contributions. Each contribution comes from a string of $\sim\!\xi_f$ grains. If fluctuations are Gaussian, with variance $\sim 1$ in our units, then we estimate $\langle \Gammab \rangle \sim M^{-1} \xi_f ( \sqrt{M}/\xi_f)^{-1/2}$.

On averaging, we find
\eq{
\langle \nabla \cdot \rhob \rangle - \nabla \cdot \rhob & = \langle \Gammab \rangle : \nabla  \rhob + \OO(\nabla \nabla \rhob).
}
In the continuum limit, if $\nabla \sim L^{-1}$ then the relative error in $(\nabla \cdot \rhob)^g$ vanishes with $1/M$ and $1/L$. Discrete calculus allows us to identify which terms in discrete expressions remain in the continuum limit.




By similar manipulations, discussed in the Appendix, we obtain a discrete calculus expression for the sum of \eqref{consist} around loops. We find
\eq{
0 & = (\nabla \cdot \rhob)^g, \label{graineqn} \\
0 & = (\Delta \varphi)^\ell + \left( \nabla \cdot \left( (\nabla \rhob) \times \rb \right)\right)^\ell + \left( \nabla (\rb \times \rhob) \right)^\ell \label{loopeqn},
}
which, together, are the exact discrete calculus reformulation of \eqref{consist}. To establish the relationship between $\sigmab$, $\varphi$, and $\rhob$, we introduce auxiliary variables 
\eq{ \label{psidef}
\psi^c_\ell \equiv \varphi^\ell - \tb^c_\ell \times \rhob^\ell, 
}
which, we will see, are the values of the Airy stress function. If we sum \eqref{psidef} around a loop, we see that 
\eq{ \label{varphi}
\varphi^\ell = \frac{1}{z^\ell} \sum_{c \in C^\ell} \psi^c_\ell, 
}
provided $\rb^\ell = \frac{1}{z^\ell} \sum_{c \in C^\ell} \rb^c$, with $z^\ell$ the number of contacts around a loop. Hence $\varphi$ is nothing but a loop average of $\psi$. Again summing \eqref{psidef} around a loop, but now weighting the sum with coefficients $\lb^c$, we find
\eq{ \label{rhob}
\rhob^\ell & = (\gb^\ell)^{-1} \cdot (\nabla \times \psi)^\ell
}
where 
\eq{ \label{gbdef}
\gb^\ell  = \frac{1}{A^\ell} \sum_{c \in C^\ell} \lb^c \rb^c \cdot \epsb = \epsb \cdot (\nabla \rb)^\ell \cdot \epsb^T,
}
is another fabric tensor. The second equality in \eqref{gbdef} indicates that $\gb$ converges to $\epsb \cdot \nabla \rb \cdot \epsb^T =\epsb \cdot \delb \cdot \epsb^T = \delb$ in the continuum limit, with $\delb$ the identity tensor. Finally, the stress tensor $\sigmab$ can be written \cite{DeGiuliMcElwaine:2011}
\eq{ \label{sigmag}
\sigmab^g = (\nabla \times \rhob)^g.
}
Hence, given the values of $\psi$ one can determine $\varphi, \rhob,$ and $\sigmab$. These relations indicate that stress corresponds to curvature of the Airy stress function $\psi$. 

The definition \eqref{psidef} has a simple geometric interpretation (Figure \ref{fig:psi}). In $(\rb,\psi)$ space the plane with normal $(\epsb \cdot \rhob^\ell,+1)$ which passes through $(\rb^\ell,\varphi^\ell)$ is described by the equation $0=(\rb-\rb^\ell,\psi-\varphi^\ell) \cdot (\epsb \cdot \rhob^\ell,+1) = (\rb-\rb^\ell) \times \rhob^\ell + \psi - \varphi^\ell$. Recalling that $\tb^c_\ell = \rb^c - \rb^\ell$, the definition \eqref{psidef} says that for each loop, we create a facet of a surface which passes through $(\rb^c,\psi^c_\ell)$ for each contact. The consistency equations \eqref{consist} are equivalent to continuity of $\psi^c_\ell$ at a contact, $\psi^c_{\ell'(c)} = \psi^c_{\ell(c)}$. Continuity of $\psi$ across a contact is equivalent to continuity of the surface at that contact. This defines a piecewise linear surface with holes at each grain. The latter can always be filled in smoothly. Hence, in the continuum, $\psi$ is a continuous function. 

The introduction of $\psi$ already indicates the physical meaning of the stress-geometry equation. Indeed, given a smooth function $\psi(\rb)$ in the continuum, the necessary and sufficient condition that $\psi^c=\psi(\rb^c)$ yields a discrete Airy stress function satisfying \eqref{psidef} for some $\rhob$ and $\varphi$ is that $\psi$ varies linearly across voids. Since stress corresponds to curvature of $\psi$, this is equivalent to requiring that stress is concentrated on the grains. This observation, made on the exact equations \eqref{graineqn} and \eqref{loopeqn}, will motivate approximations needed below to obtain non-trivial continuum equations. 

\section{Mean-field}
 

Each average defined by \eqref{average} can be considered an expectation value over the quenched geometry; terms involving products then involve correlations. The discrete calculus formulation exactly accounts for correlations on the scale of a single grain and a single loop. As above, we assume that the geometry is homogeneous on a mesoscopic scale $\xi_f$ with $1 \ll \xi_f \ll L$. We then expect the mesoscopic-scale correlations to be small. Neglecting them, we assume $\langle \rb \rhob \rangle = \langle \rb \rangle \langle \rhob \rangle$ and $\langle \rhob \rangle = \langle \gb \rangle^{-1} \cdot \langle \nabla \times \psi \rangle$. The success of mean-field theories of frictionless isostatic packings supports this assumption (see, e.g. \cite{Wyart:2010}).

Identifying averaged quantities with continuum ones and immediately dropping the $\langle \cdot \rangle$ decoration, the preceding equations \eqref{graineqn}, \eqref{loopeqn}, \eqref{rhob}, \eqref{gbdef}, and \eqref{sigmag} then give continuum equations $\gb=\delb$ and
\eq{
0 & = \nabla \cdot \rhob \label{divrho2} \\
0 & = \Delta (\varphi - \psi) \label{phipsi} \\
\rhob & = \nabla \times \psi \\
\sigmab & = \nabla \times \nabla \times \psi. \label{sigma2}
}
Because $\rhob = \nabla \times \psi$, the continuum equation $\nabla \cdot \rhob=0$ is identically satisfied. We conclude that the continuum stress geometry equation is $\Delta (\varphi-\psi)=0$. It remains to establish the continuum relation between $\varphi$ and $\psi$. The discrete relation \eqref{varphi} suggests the naive closure $\varphi=\psi$; however, this would give a continuum equation which is identically satisfied. As we noted from examining the discrete equations, the stress-geometry equation imposes concentration of stress on the grains. To understand why this effect leads to nontrivial $\psi - \varphi$, we consider the geometric interpretation of $\psi$. 

The discrete Airy stress function $\psi$ describes a continuous, but patchwork surface, which is alternately flat and curved on voids and grains, respectively. In cross-section, this surface appears as the solid curve in Figure \ref{fig:cross}. In the continuum, grains and voids are not well-defined: the effect of averaging is to replace the original patchwork surface with a coarse-grained surface $\langle \psi \rangle$ which is \textit{not} flat on voids, depicted by the dashed curve in Figure \ref{fig:cross}. Each loop shrinks to a point, and the loop equation \eqref{loopeqn} becomes an equation valid at the point $\rb^\ell$. The equation $\Delta (\varphi-\psi)=0$ thus relates $\varphi(\rb^\ell)=\langle \varphi \rangle (\rb^\ell)$ to $\psi(\rb^\ell)=\langle \psi \rangle (\rb^\ell)$. 
Crucially, because contact forces are repulsive, the $\langle \psi \rangle$ surface has positive mean curvature; hence $\psi(\rb^\ell)\equiv \langle \psi \rangle(\rb^\ell)$ will \textit{systematically} deviate from $\varphi^\ell$. 

The average $\langle \psi \rangle(\rb^\ell)$ is not defined by \eqref{average}, but since the coarse-grained surface is assumed smooth, it will suffice to use Taylor expansion.
By homogeneity, it is reasonable to force $\langle \psi \rangle(\rb)$ to go through all $\psi^c$. Then, we can simply Taylor expand $\langle \psi \rangle(\rb^c)$ around $\langle \psi \rangle(\rb^\ell)$, and compute $\varphi^\ell=\ffrac{1}{z^\ell} \sum_{c \in C^\ell} \langle \psi \rangle(\rb^c)$ exactly, introducing fabric tensors which characterize the local geometry. Here we will fit $\langle \psi \rangle(\rb)$ to a quadratic polynomial around a loop; higher-order terms are suppressed by powers of $1/L \ll 1$.
We have 
\eq{
\langle \psi \rangle(\rb) & = \langle \psi \rangle(\rb^\ell) + \bm{h} \cdot \nabla \langle \psi \rangle(\rb^\ell) + \half  \bm{h}\bm{h} : \nabla \nabla \langle \psi \rangle(\rb^\ell) \notag 
}
with $\bm{h} = \rb - \rb^\ell$ and hence
\eq{
\varphi^\ell & = \langle \psi \rangle(\rb^\ell) + \half \Fb^\ell : \nabla \nabla \langle \psi \rangle(\rb^\ell),
}
defining a fabric tensor
\eq{
\Fb^\ell = \frac{1}{z^\ell} \sum_{c \in C^\ell} \tb^c_\ell \tb^c_\ell. 
}
Through its principal axes and eigenvectors, $\Fb^\ell$ can be physically associated with an ellipse which fits the loop $\ell$. Its average
\eq{
\Fb(\rb) = \frac{1}{2N_C(L)} \sum_{\ell \in L} \sum_{c \in C^\ell} \tb^c_\ell \tb^c_\ell
}
defines the ellipse which best fits loops around $\rb$; higher-order fabric tensors would measure refinements of this shape. With these definitions, we expect $\varphi - \psi = \half \Fb : \nabla \nabla \psi$ in the continuum and hence
\eq{ \label{eqn!}
0 & =  \Delta \left( \bm{\hat{F}} : \nabla \nabla \psi \right), 
}
which is the continuum mean-field stress-geometry equation, to leading order in $1/L$. It can be written in terms of $\sigmab$ directly using $\nabla \nabla \psi = \epsb^T \cdot \sigmab \cdot \epsb = \tr(\sigmab) \delb - \sigmab$.

As discussed above, discrete calculus ignores fluctuating error terms which exist for any finite $M$. The leading error in \eqref{eqn!} can be as large as the error in \eqref{divrho}, which is $\Gammab : \nabla \rhob\sim L^{-2} M^{-5/4} \xi_f^{3/2} \psi$; this must be much smaller than the term in \eqref{eqn!}, which indicates that the continuum theory can only hold when $\xi_f \ll M^{5/6} L^{-4/3}$. Since a continuum theory can only resolve stress gradients much larger than the averaging scale, we must also have $M^{1/2} \ll L$. This implies that we must take $M \gg \xi_f^6$ independent of $L$. In particular, if the fabric correlation length 
$\xi_f \to \infty$, then the continuum theory breaks down and noise will dominate the response. The sensitive dependence on $\xi_f$ may explain the noisy response observed in many granular materials. Mathematically, this arises because the leading order terms in \eqref{phipsi} cancel; physically, this is because stress localization onto grains is a microscopic property.


\section{Discussion} \subsection{Hyperbolic vs. elliptic} 
The mathematical form of \eqref{eqn!} depends on the fabric tensor $\Fb$; where the latter has strong spatial gradients, \eqref{eqn!} is nearly hyperbolic, and where $\Fb$ is spatially homogeneous, \eqref{eqn!} is elliptic\footnote{Note that equation \eqref{eqn!} can be put into the form of \eqref{eqseries} by letting $\Fb_2=0, \Fb_4 = \delb\delb \tr(\Fb)-\delb\Fb$, and using the identity $\epsb^T \cdot \bm{\hat{A}} \cdot \epsb = \delb \tr(\bm{\hat{A}})-\bm{\hat{A}}^T$. }. To illustrate this, suppose $\Fb(\rb)=\Fb_0+\Fb_1 h(\rb)$, where\footnote{Explicitly, $||\bm{\hat{A}}||\equiv\sqrt{\sum_{i,j} A_{ij} A_{ji}}$.} $||\Fb_1|| \ll ||\Fb_0||$ but $h$ fluctuates on the small scale $
\xi_f$. If we linearize \eqref{eqn!} around $\rb=0$, where for simplicity we assume $h(0)=\nabla h(0)=0$, then \eqref{eqn!} becomes $0=(\Delta h) \Fb_1 : \nabla \nabla \psi + \Fb_0 : \nabla \nabla \Delta \psi$. Seeking solutions $\psi \propto e^{i \bm{k}\cdot \rb}$ leads to the `dispersion relation' $(\Delta h) \bm{k}\cdot \Fb_1 \cdot \bm{k} = k^2 \bm{k}\cdot \Fb_0 \cdot \bm{k}$. Since $\Fb_0$ defines the mean shape of loops, it has positive eigenvalues, and hence when $\Fb_1=0$, the dispersion relation has no solutions for \textit{real} $\bm{k}\neq 0$. This means that all solutions have a decaying component and the response is elliptic. However, when $\Fb_1 \neq 0$, then we can have solutions for real $\bm{k}$ if $k^2 \sim \Delta h ||\Fb_1||/||\Fb_0|| \sim \xi_f^{-2} ||\Fb_1||/||\Fb_0||$, giving purely oscillatory (hyperbolic) solutions.  Simulations suggest that the fabric varies smoothly \cite{RoulSchinner:2010}, so that $||\Fb_1||/||\Fb_0||$ should go to zero with $1/M$, and hence elliptic behaviour is generically expected at the macroscopic scales where the theory is applicable.
  
\subsection{Fabric} The fabric tensor $\Fb$ contains information about packing inhomogeneity and anisotropy, as expressed through the size and shape of loops. Its trace is approximately $\xi^2 \equiv \tr(\Fb) \sim ((\zbar-2) \phi (1-x_0))^{-1}$, with $\phi$ the area fraction and $x_0$ the fraction of `rattlers,' grains which are trapped in the packing but do not contribute to mechanical stability. 
The dominant geometrical dependence is through $\zbar -2$, which may vary by a factor of two over a packing\footnote{For rigid grains, over the realistic range $0.78 < \phi < 0.84$, $0 < x_0 < 0.15$, and $3 < \zbar < 4.5$, $\phi$, $x_0$, and $\zbar-2$ vary by factors of $1.08, 1.18,$ and 2.5, respectively.}. $\Fb$ also describes anisotropy in the contact distribution, which develops under shear \cite{Oda:1972a,Oda:1972b,RadjaiRoux:2005,Radjaietal:2012}. In terms of the more frequently used fabric tensor $\bm{\hat{F}}_C(\rb) = \frac{1}{N_{C}(C)} \sum_{c \in C} \lb^c \lb^c$, we have approximately $\Fb \approx \xi^2 \; \epsb \cdot \bm{\hat{F}}_C \cdot \epsb^T$, implying that $\Fb$ and $\Fb_C$ share principal axes. 

In the simplest isotropic and homogeneous case $\Fb(\rb) = \half \xi^2 \delb$, the stress-geometry equation reduces to the biharmonic equation $\Delta \Delta \psi = 0$, which is the same equation satisfied by the Airy stress function $\psi$ in isotropic elasticity \cite{Muskhelishvili}. Note that $\psi$ corresponds to the \textit{total} stress, as opposed to a stress increment, as discussed below.  It is noteworthy that we derive this result without reference to strain, Hooke's law, or energy. It explains the success of isotropic elasticity in the presence of an isotropic fabric \cite{Bouchaudetal:2001, Sereroetal:2001,Atmanetal:2005}. 

More generally, a homogeneous but anisotropic fabric yields the equation described by $\psi$ in anisotropic elasticity \cite{Sadd:2009}. Anisotropy induces stretching and rotation of stress contours, as observed in experiments \cite{Sereroetal:2001,Gengetal:2001b,Gengetal:2003,Atmanetal:2005}. 
In the special case of stress-only boundary conditions and a homogeneous fabric, the stress-geometry equation thus recovers anisotropic elasticity. To apply boundary conditions on displacements would require 
an analog of Hooke's law for \textit{rigid grains}, so far absent \cite{GaydaSilveira:2007}. 


\subsection{Friction}
The result \eqref{eqn!} was derived assuming rigid, perfectly frictional grains at isostaticity, in the absence of gravity. However, none of these assumptions were essential. 
Here we discuss the effect of finite friction. The effects of finite stiffness, hyperstaticity, and gravity will be discussed elsewhere (see \cite{DeGiuli:2013}).


Friction strongly constrains solutions to \eqref{eqn!}. Each contact force must satisfy the Coulomb friction inequality $|f_T^c| \leq \mu_f f_N^c$, where $f_T^c$ and $f_N^c$ are the tangential and normal components of the contact force at $c$, and $\mu_f$ is the Coulomb friction coefficient. In 2D, this can be written as the pair of inequalities $( \Gb^c \cdot \bm{\hat{M}}_\pm ) : \sigmab^c \geq 0$, where $\Gb^c = \bm{n}^c \bm{n}^c$ is a fabric tensor constructed from contact normals $\bm{n}^c$, and $\bm{\hat{M}}_\pm = \delb \pm \ffrac{1}{\mu_f} \epsb$. Under the same mean-field assumptions as earlier, this yields the pair of continuum inequalities
\eq{ \label{Coulomb}
\big( \Gb \cdot \bm{\hat{M}}_\pm \big) : \sigmab & \geq 0.
}
Summing these inequalities implies $\Gb : \sigmab \geq 0$, a generalization of positive pressure $P \geq 0$. In the frictionless limit $\mu_f \to 0$, \eqref{Coulomb} implies $(\Gb \cdot \epsb) : \sigmab = 0$, which states that $\Gb$ and $\sigmab$ share principal axes. Note that $\Gb=\Fb_C$ for disks.

The stress-geometry equation needs to be solved subject to the Coulomb inequalities \eqref{Coulomb}\footnote{When the contact normal distribution is sharply peaked about a pair of perpendicular directions $\bm{n}$ and $\epsb \cdot \bm{n}$, this reduces to the Mohr-Coulomb inequality $|\sigma_{nt}| \leq \mu \; \sigma_{nn}$.}. If a prospective solution violates one of these inequalities, then failure must occur within the material; the location of failed regions must be tracked by dynamical evolution of the fabric, beyond the scope of the present theory.

\subsection{Stress increments}

Equation \eqref{eqn!} governs the \textit{total} stress $\sigmab$, which has important consequences for stress increments. Consider a small deformation of strain $e\!\ll\!1$ which takes the packing from $(\Fb,\psi)$ to $(\Fb+\delta \Fb, \psi+\delta \psi)$ due to some change in boundary conditions. Even for rigid grains, the change in fabric $\delta \Fb$ can be nonnegligible, due to contact opening, closing, and sliding. Suppose $\delta \Fb\! \sim\!e$ but is otherwise unknown. Then by ignoring $\delta \Fb$, and given the initial fabric, the initial and final stress can be determined up to a relative error which goes to zero with $e$. Up to some larger relative error, $\delta \psi$ is then also known. But taking the difference of \eqref{eqn!} in initial and final states gives, to linear order, $0 = \Delta(\Fb : \nabla \nabla \delta \psi + \delta \Fb : \nabla \nabla \psi)$, indicating that $\delta \psi$ depends on $\delta \Fb$. The size of the error in $\delta \psi$ is related to the size of $\delta \Fb$. This counterintuitive behaviour is a consequence of isostaticity: knowledge of fabric implies knowledge of stress.



%

\section{Conclusion}

To summarize, in this work we have derived the missing stress-geometry equation for 2D frictional isostatic granular materials, equation \eqref{eqn!}, in the mean-field approximation. The equation imposes that the voids cannot carry any stress. Formally, the equation resembles the stress form of St. Venant's compatibility condition in anisotropic elasticity, but (i) it governs the total stress $\sigmab$, as opposed to stress increments, and (ii) it can be derived without reference to strain. It must be solved subject to the (continuum) Coulomb inequalities \eqref{Coulomb}, which are necessary for local mechanical stability. 
The theory emphasizes the need to understand fabric evolution.

That so much can be said without mention of strain is a consequence of realistic granular materials being nearly rigid and close to isostaticity. This \textit{does not} imply that strain cannot be defined for such a material; indeed, strain can easily be defined based on the deformation of loops \cite{Satake:1993,KruytRothenburg:1996,DeGiuli:2013}. In order to apply boundary conditions on displacement, and understand fabric evolution, such a variable would be necessary to complete the present theory; we leave this for future work \cite{DeGiuli:2013}.

To derive \eqref{eqn!} we have assumed that (i) the geometry and the stress are uncorrelated on a mesoscopic scale $\xi_f \gg 1$, (ii) macroscopic stress gradients occur on a still larger scale $L \gg \xi_f$, (iii) continuum variables are defined by averaging over a region of linear size $M^{1/2} \gg \xi_f^3$, and (iii) grains can be treated as rigid. 

Subject to these assumptions, and with knowledge of the fabric, the present theory can be used to solve for the stresses inside a stable granular solid with boundary conditions on stress. What happens when a Coulomb inequality is saturated, how the fabric evolves, and how displacement boundary conditions can be applied, are questions left for future work. 


%
%
%
%

\acknowledgments
E.D. gratefully acknowledges illuminating discussions with N.J. Balmforth, R. Blumenfeld, J.P. Bouchaud, O. Dauchot, I. Hewitt, J. McElwaine, and F. Radjai, and NSERC for funding.


\section{Appendix. Discrete calculus} 
\renewcommand{\theequation}{A.\arabic{equation}}
\setcounter{equation}{0}

In the main text, we defined $(\nabla \cdot \rhob)^g$ as a weighted sum of terms around the grain $g$, with the weights determined by Stokes' Theorem. All other discrete derivatives can be defined similarly \cite{Schwalmetal:1999,DeGiuliMcElwaine:2011,DeGiuli:2013}. For example, Stokes' Theorem motivates a definition for a loop divergence $A^\ell (\nabla \cdot \bm{F})^\ell \equiv -\sum_{c \in C^\ell} \lb^c \times \bm{F}^c$ for a field $\bm{F}$ defined on contacts. In the text, we apply this to $\bm{F}^c = (\nabla \rhob)^c \times \rb^c$, requiring a definition of a discrete gradient. The gradient theorem motivates $A^c (\nabla \rhob)^c = \epsb \cdot \lb^c \left( \rhob^{\ell'} - \rhob^\ell \right),$ where $A^c = (\bm{t}^c_{\ell'}-\bm{t}^c_\ell) \times \lb^c$ is a signed area associated to $c$. These definitions imply a discrete Laplacian $A^\ell (\Delta \varphi)^\ell \equiv \sum_{c \in C^\ell} (|\lb^c|^2/A^c)  \left( \varphi^{\ell'} - \varphi^\ell \right).$ Finally, the natural generalization to discrete loop and grain gradients $A^\ell (\nabla \bm{F})^\ell \equiv -\sum_{c \in C^\ell} \lb^c \cdot \epsb \; \bm{F}^c$ and $A^g (\nabla \rhob)^g \equiv -\sum_{\ell \in L^g} \sb^\ell_g \cdot \epsb \; \rhob^\ell$, together with $\nabla \times \bm{F} = \epsb \cdot \nabla \bm{F}$, valid in the plane, complete the needed definitions. 

Writing $\tb^c_\ell = \rb^c-\rb^\ell$, summing \eqref{consist} around a loop, and straightforwardly applying these definitions leads to \eqref{loopeqn} in the main text.

\bibliographystyle{eplbib}
\bibliography{../me/Granular-2-1}

%
%


\end{document}